\documentstyle[12pt,aasms]{article}
\articleid{11}{14}
\begin{document}

\title{ NON-LINEAR APPROXIMATIONS TO GRAVITATIONAL INSTABILITY:
A COMPARISON IN SECOND-ORDER PERTURBATION THEORY}

\author {Dipak Munshi}
\affil {Inter-University Center for Astronomy and Astrophysics
Post Bag 4, Ganeshkhind, Pune 411007, India \\ munshi@iucaa.ernet.in }
\and
\author {Alexei A. Starobinsky}
\affil {Landau Institute for Theoretical Physics
Kosygina St. 2, Moscow 117334, Russia \\ alstar@cpd.landau.free.msk.su}

\begin{abstract}

Nonlinear approximation methods such as the Zeldovich approximation, and
more recently the frozen flow and linear potential approximations, are
sometimes used to simulate nonlinear gravitational instability
in the expanding Universe. We investigate the relative accuracy of these
approximations by comparing them with the exact solution using  second order
perturbation theory. We evaluate the density and velocity fields in these
approximations to second order, and
 also determine the skewness parameter $S_3 = \langle \delta^3
\rangle / \langle (\delta^{(1)})^2\rangle ^2$ for each of the approximations
again in second order. We find
that $S_3 = 4, ~3, ~3.4$ for the Zeldovich
approximation, the frozen flow and the linear potential approximations
respectively as compared to $S_3 = {34/7}$ for the exact solution.
Our results show that, of all the approximations considered, the Zeldovich
approximation is the most
accurate in describing the weakly nonlinear effects of gravity. Moreover, the
Zeldovich approximation is much closer to the exact results for matter
and velocity distributions than the
other approximations if the slope of the power spectrum of density
perturbations is $-3< n \le -1$.
\end {abstract}

\keywords{Cosmology, gravitational clustering, nonlinear
approximations, non-Gaussian statistics.}

\section{Introduction}
It is now commonly agreed that all gravitationally bound objects in the
Universe as well as its large-scale structure originated from the growth of
initially small inhomogeneities in the expanding Universe. Generally
speaking the
investigation of non-linear evolution of non-interacting particles
constituting the dark matter in the Universe requires long N-body
computer simulations and as a result has been performed only for a
selected subclass of initial conditions.
That is one of the reasons why different approximation schemes have
been proposed which
greatly reduce the required numerical work and make it possible to
investigate non-linear evolution for a much wider class of initial
conditions and for longer periods of time.

Amongst nonlinear approximation methods the following three are the
most natural and straightforward: \\
1) the Zeldovich approximation (Zeldovich 1970) and its further
generalization to the period after caustics formation - the adhesion
model (Gurbatov, Saichev \& Shandarin 1985); \\
2) the frozen flow approximation (Matarrese et al. 1992); \\
3) the linear potential approximation (Brainerd et al. 1992;
Bagla \& Padmanabhan 1993). \\
All these approximations are really {\it approximations} in the sense
that they are neither exact nor asymptotic to the exact solution beyond
linear order (apart from some special degenerate cases).
The difference between each of these approximations and the exact solution
arises
already in the second order of perturbation expansion. Thus, a natural
way to see the difference between these approximations and to estimate
their relative accuracy is to calculate
their departure from the exact solution in this order. This is just the
aim of the present paper.

\section{Second Order Perturbative Calculations}
Let us consider gravitational instability in the spatially flat matter
dominated
FRW Universe before the formation of caustics. In such a
Universe  the scale factor grows as $a(t)\propto t^{2/3}$ and the
background matter density decreases according to $\rho_0=1/6\pi Gt^2$. The
equations describing this process
have the following form in the Newtonian approximation ( see, e.g.
Peebles 1980):

\begin{eqnarray}
\triangle \Phi =4\pi Ga^2\rho_0\delta;~~~~ \\
\dot \delta + {1\over a} div((1+\delta)\vec u)=0; \\
(a\vec u)^{\cdot}+(\vec u \nabla)\vec u = -\nabla \Phi~~
\end{eqnarray}
where $\delta ={(\rho -\rho_0)/ \rho_o}$.

The velocity $\vec u$ is irrotational, therefore it is possible to introduce
a velocity potential $V$ so that $\vec u= -\nabla V/a$. Then Eq.
(3) may be substituted by its first integral:

\begin{equation}
\dot V= \Phi +{1\over 2a^2}(\nabla V)^2,
\end{equation}

\noindent
and the first two equations acquire the form:

\begin{eqnarray}
\triangle \Phi= {2a^2\over 3t^2}\delta;~~~~~~~~\\
a^2\dot \delta= (1+\delta)\triangle V+ \nabla \delta \nabla V.
\end{eqnarray}

Let us expand all quantities into series in powers of an initial
density enhancement: $\delta=\delta^{(1)}+\delta^{(2)}+...$, and the
same for $\Phi$ and $V$. As a function of time, this is an expansion in
powers of $t^2/a^2\propto t^{2/3}$ (since $\delta^{(n)} \propto
t^{{2n\over 3}}$). The first order solution (the linear approximation)
is the same for all three abovementioned approximation schemes:

\begin{eqnarray}
\Phi^{(1)}=\phi_0;~~~ \,\,
V^{(1)}=\phi_0t;~~~ \,\,
\delta^{(1)}={3t^2\over 2a^2}\triangle \phi_0
\end{eqnarray}

\noindent
where $\phi_0(\vec r)$ is the initial gravitational potential.
The value and statistical properties  of $\phi_0(\vec r)$
are completely arbitrary for classical cosmology
(apart from the trivial condition of smallness: $|\phi_0(\vec r)|\ll
1$). On the other hand, any theory of the Early Universe should
produce some predictions for the properties of $\phi_0$. In particular,
the inflationary scenario of the Early Universe predicts that $\phi_0$
is a gaussian stochastic quantity with zero average and dispersion
for the Fourier transform $\phi_0(k)$
having an approximately $k^{-3}$ dependence (in
the simplest versions of this scenario and before the multiplication of
$\phi_0(k)$ by a transfer function).

The form of the linear approximation (7) naturally leads to
three approximation schemes which arise as a result of imposing by hand some of
the relations valid in this approximation on a fully non-linear solution. Of
course, having introduced one new relation, one has to abandon one of the
previous equations. That abandoned equation is chosen to be the Poisson
equation
(1) or (5). Thus, in all these approximations we neglect the self-gravity
of inhomogeneities. As a result, we have the following relations for the
full series: \\
1)$V=\Phi t$ - the Zeldovich approximation (further denoted by ZA and by the
subscript $Z$); \\
2)$V=\phi_0 t$ - the frozen flow approximation (FF, subscript $f$); \\
3)$\Phi=\phi_0$ - the linear potential approximation (LP, subscript $p$). \\
Our way of introducing the Zeldovich approximation is different from that
commonly
found in the literature. However, it is straightforward to check that the
formula
$V=\Phi t$ provides both the necessary and sufficient conditions for
conventional
expressions of the Zeldovich approximation, since, (a) it can directly be
derived   from those latter expressions, and (b) by inserting $V=\Phi t$
 in Eq. (4) and then solving Eq. (4)
using a trivial change of variables $( V = a^2\dot a V')$, we arrive  at the
Zeldovich - Bernoullie equation  commonly used to describe the evolution of the
velocity potential in the Zeldovich approximations (see eg. Kofman 1991) which
is given as follows
$${\partial \over \partial a} V' + {1\over2} ({\nabla V'})^2 = 0.$$
Let us now consider the second order terms. \\

\medskip\noindent
1) The Exact second order solution

\begin{eqnarray}
\triangle \Phi^{(2)}={2a^2\over 3t^2}\delta^{(2)};~~~~~~~~~~~~~~ \\
a^2\dot \delta^{(2)}=\triangle V^{(2)}+ \delta^{(1)}\triangle V^{(1)}
+\nabla \delta^{(1)}\nabla V^{(1)}; \\
\dot V^{(2)}=\Phi^{(2)}+{1\over 2a^2}(\nabla V^{(1)})^2.~~~~~~~~~
\end{eqnarray}

\noindent
After excluding $\Phi^{(2)}$ and $V^{(2)}$ from these equations, we get
an equation for $\delta^{(2)}$:

\begin{equation}
\ddot \delta^{(2)}+{4\over 3t}\dot \delta^{(2)}-{2\over
3t^2}\delta^{(2)}= {1\over a^2}{\partial \over \partial t}
(\delta^{(1)}\triangle V^{(1)}+
\nabla \delta^{(1)} \nabla V^{(1)}) +{1\over 2a^4}\triangle
\left( (\nabla V^{(1)})^2\right).
\end{equation}

\noindent
Note that the left hand side of this equation is the same as in the first
order. The solution of (8-11) is (we consider the growing mode only):

\begin{eqnarray}
\delta^{(2)}={9t^4\over 28a^4} (5P+\triangle Q);~~~~~~~~~~~~~~\nonumber \\
\Phi^{(2)}={3t^2\over 14a^2} (5\triangle^{-1}P+Q);~~~~~~~~~~ \nonumber \\
V^{(2)}={3t^3\over 7a^2}({3\over 2}\triangle^{-1}P+Q);~~~~~~~~~~~ \nonumber \\
P(\vec r)=(\triangle \phi_0)^2+\nabla \phi_0 \nabla (\triangle \phi_0)
=\nabla (\nabla \phi_0 \triangle \phi_0),\nonumber \\
Q(\vec r)=(\nabla \phi_0)^2.~~~~~~~~~~~~~~~~~~
\end{eqnarray}

\noindent
Of course, this solution is well known (see, e.g. Peebles 1980).
The average value of $\delta^{(2)}$ is zero because it has the form of
divergence (the same is true in all orders). The expression for
$\delta^{(2)}$ is local while the appearance of the inverse
Laplacian $\triangle^{-1}$ in the
expressions for $\Phi^{(2)}$ and $V^{(2)}$ shows that they are
non-local, as is the velocity $\vec u^{(2)}$ (but $div \, \vec
u^{(2)}$ is local).  Non-local terms in the expansion of $\delta$
begin from $\delta^{(3)}$, thus, the first nonlinear corrections to the
power spectrum and the density correlation function are non-local, too.

\medskip\noindent
2) The Zeldovich approximation

\begin{eqnarray}
\Phi_Z^{(2)}={V_Z^{(2)}\over t}; ~~~~\,
\dot V_Z^{(2)}=\Phi_Z^{(2)}+{1\over 2a^2}(\nabla V^{(1)})^2,
\end{eqnarray}

\noindent
the third equation is the same as Eq.(9). The solution is:

\begin{eqnarray}
\delta_Z^{(2)}={9t^4\over 16a^4}(2P+\triangle Q); ~~~\,
V_Z^{(2)}=\Phi_Z^{(2)}t={3t^3\over 4a^2}Q.~~
\end{eqnarray}

Note that in the case of one-dimensional plane-symmetric motion $\phi_0
= \phi_0(x)$, $P= {1\over 2}\triangle Q  $ and
the above terms coincide with the second-order terms for the
exact solution (12). This is a consequence of the fact that the
Zeldovich approximation is actually an exact solution of Eqs. (4 - 6)
in the case of one-dimensional motion before caustic formation (Shandarin \&
Zeldovich 1989). \\

\medskip\noindent
3) The frozen flow approximation

\begin{equation}
V_f^{(2)}=0; \; \Phi_f^{(2)}=-{1\over 2a^2}(\nabla V^{(1)})^2=
-{t^2\over 2a^2}Q,
\end{equation}

\noindent
the third equation being the same as Eq.(9). The only quantity that
remains to be found is $\delta_f^{(2)}$. From the above equations it
follows that:

\begin{equation}
\delta_f^{(2)}={9t^4\over 8a^4}P.
\end{equation}

\medskip\noindent
4) The linear potential approximation

\begin{equation}
\Phi_p^{(2)}=0; \; \dot V_p^{(2)}={1\over 2a^2}(\nabla V^{(1)})^2,
\end{equation}

\noindent
the third equation still being the same as Eq.(9). Solutions for the
remaining quantities are

\begin{equation}
V_p^{(2)}={3t^3\over 10a^2}Q; \; \delta_p^{(2)}=
{9t^4\over 40a^4}(5P+\triangle Q).
\end{equation}

\noindent
Note that $\delta_p^{(2)}=0.7\delta^{(2)}$. Thus, the
second order correction in the linear potential approximation has the
same spatial structure as the exact solution but its value is
$30\%$ smaller than that of the exact solution.

\clearpage
\section{Comparison of Approximations}
\subsection{Density Perturbations}
Comparing second order terms in the density perturbation in the different
approximation schemes, we see that the P terms are the same in all of them (and
different from the exact solution). The
difference between $ \delta_Z^{(2)}, \delta_f^{(2)}$ and $\delta_p^{(2)}$
arises essentially because of the different numerical coefficients in
front of Q in each of these approximations.
In addition we would like to point out that there are no nonlocal terms
in $\delta^{(2)}$ in any of these approximations.

Let us now consider the difference $\Delta$ between the approximate solutions
and the exact
one in the second order,  and also calculate the expected variances of $\Delta$
assuming that the initial potential $\phi_0(\vec r)$ is a gaussian
stochastic quantity with zero average and an isotropic power spectrum.
We have

\begin{eqnarray}
\Delta_Z\equiv \delta_Z^{(2)} -\delta^{(2)}=-{27t^4\over 56a^4}
(P-{1\over 2}\triangle Q); \nonumber ~~~~~~~\\
\Delta_f\equiv \delta_f^{(2)} -\delta^{(2)}=-{27t^4\over 56a^4}
(P+{2\over 3}\triangle Q); \nonumber ~~~~~~~ \\
\Delta_p\equiv \delta_p^{(2)} -\delta^{(2)}=-{27t^4\over 56a^4}
(P+{1\over 5}\triangle Q)=-0.3\delta^{(2)}.
\end{eqnarray}

We introduce the notations: $\sigma_1^2=\langle (\nabla
\phi_0)^2\rangle$; $\sigma_2^2=\langle (\triangle \phi_0)^2\rangle$;
$\sigma_3^2=\langle (\nabla(\triangle \phi_o))^2\rangle$. In the linear
approximation, $\sigma_1^2$ is proportional to the velocity dispersion
($\sigma_{v}^2\equiv \langle (\vec u^{(1)})^{2}\rangle =
{t^2\over a^2}\sigma_1^2$),
$\sigma_2^2$ - to the dispersion of density perturbations
($\sigma_{\rho}^2\equiv \langle (\delta^{(1)})^{2}\rangle =
{9t^4\over 4a^4}\sigma_2^2$).
$\sigma_2^2\le \sigma_1 \sigma_3$ with the equality being achieved
in the case
when the Fourier spectrum is proportional to $\delta (k-k_0)$ only.
Using the useful relations

\begin{eqnarray}
\langle P^2\rangle = {7\over 3}\sigma_2^4 + {1\over 3}\sigma_1^2
\sigma_3^2; \nonumber \\
\langle P\triangle Q\rangle =  2\sigma_2^4 + {2\over 3}\sigma_1^2
\sigma_3^2; \nonumber~ \\
\langle (\triangle Q)^2\rangle = {44\over 15}\sigma_2^4 + {4\over 3}
\sigma_1^2 \sigma_3^2,
\end{eqnarray}
after lengthy but straightforward calculations we get
\begin{eqnarray}
\langle \Delta_Z^2\rangle={16\over 15}\sigma_2^4A^2;  \nonumber ~~~~~~~~~~\\
\langle \Delta_f^2\rangle=\left({851\over 135}\sigma_2^4+
{49\over 27}\sigma_1^2\sigma_3^2\right)A^2; \nonumber \\
\langle \Delta_p^2\rangle=\left({1219\over 375}\sigma_2^4+
{49\over 75}\sigma_1^2\sigma_3^2\right)A^2
\end{eqnarray}
where $A ={27t^4\over 56a^4}$. To obtain a relative accuracy with respect to
the second order term in the exact solution, these results should be divided
by $\langle \delta^{(2)2}\rangle = {100\over 9} \langle \Delta_p^2\rangle $.
The form of the fractional error $F_{\delta}= \langle \Delta^2\rangle/
\langle\delta^{(2)2} \rangle$  for different approximations can be expressed
as $F_{\delta}=(a\gamma^2+b)/(c\gamma^2+1)$ where $\gamma =\sigma^2_2/
{\sigma_1\sigma_3}$. The values of a, b, c for the three approximations
considered by us are (0.15, 0.0, 6.25) for ZA; (1.1, 0.25, 6.25) for FF;
and (0.0, 0.09, 0.0) for LP respectively.
{}From the plot (Fig.1) of $F_{\delta}$ as a function of $\gamma$, it is clear
that the Zeldovich
approximation is always better than either FF or LP. In
particular, for a $\delta$-like power spectrum ($\gamma =1$)
the fractional error $F_{\delta}$ is   $0.0246$, for ZA, $0.187$, for FF and
$0.090$ for LP.  In the reverse case of a very extended spectrum
$\gamma \ll 1 $ the fractional errors have the asymptotic forms
$ 0.15  \gamma^2;\, 0.25;\, 0.09 $ for ZA, FF, LP respectively. (For a
power-law
spectrum $({\delta \rho \over \rho})_{\vec k}^2 \propto k^n $, $\gamma \ll 1 $
 if the spectral index lies in the range $-5 \le n \le -1$). Note that in the
latter case the Zeldovich approximation is {\it much} closer to the exact
solution than the other two approximations since $( \langle \Delta_Z^2\rangle
\ll \langle \Delta_{f,p}^2\rangle )$.

\medskip

These results may also be used to compare the value of the skewness
parameter
$S_3=\langle \delta^3\rangle /\sigma_{\rho}^4$ which arises in each of the
above approximations with that obtained in the exact solution (Peebles 1980,
Grinstein $\&$ Wise 1987, Bouchet et al. 1992).
To first order in perturbation theory $\langle \delta^3\rangle =
\langle (\delta^{(1)})^3\rangle = 0$.
In second order, however, $\langle \delta^3 \rangle =
3\langle (\delta^{(1)})^2 \delta^{(2)}\rangle$, so that
$S_3= {34\over 7}\approx 4.86;\,4;\,3;\,3.4$ for the exact solution and for the
Zeldovich, frozen-flow and linear potential approximations respectively.
All three approximations produce a low value for the skewness but the Zeldovich
approximation is the closest to the correct answer once more ( being accurate
to
within 20\% ). It is also interesting that in all four cases the
skewness does not depend upon the form of the initial spectrum.

\subsection{Peculiar Velocities}
We now proceed to calculate the error in the peculiar velocity for
each of the approximations considered earlier.
Let $\vec U^{(2)}_Z = \vec u^{(2)}_Z - \vec u^{(2)}$ denote the difference
between the second order velocity field in the ZA and in the exact second
order analysis (similarly for FF and LP).
Let us also define $M^2 = \langle (\nabla
(\triangle^{-1} P))^2 \rangle = - \langle P\triangle^{-1} P\rangle $ (this
will be the only non-local term in the answer).
Using the useful relations

\begin{eqnarray}
\langle\nabla(\triangle^{-1}P)\nabla Q\rangle = -\langle PQ \rangle =
{2\over 3}\sigma_1^2\sigma_2^2;\nonumber \\
\langle \nabla Q \nabla Q\rangle =  {4\over 3}\sigma_1^2\sigma_2^2;
\,~~~~~~~~~~
\end{eqnarray}
we get in the second order:

\begin{eqnarray}
\langle \vec U_Z^{(2)2}\rangle = B^2(M^2 - {1\over 3}\sigma_1^2\sigma_2^2);
 \nonumber  \\
\langle \vec U_f^{(2)2}\rangle =  B^2(M^2 + {40\over 27}\sigma_1^2\sigma_2^2);
 \nonumber  \\
\langle \vec U_p^{(2)2}\rangle =  B^2(M^2 + {8\over 25}\sigma_1^2\sigma_2^2),
\end{eqnarray}
where $B = {9t^3\over 14a^3}$. In order to obtain a relative accuracy, each of
these expressions should
be divided by $\langle \vec u^{(2)2}\rangle$.
Since $\vec u_f^{(2)} = 0$, $\vec U_f^{(2)} = -\vec u^{(2)}$, therefore
this is equivalent to dividing by
$\langle \vec U_f^{(2)2}\rangle$.
{}From Eq. (23) it is clear that of the three approximations the Zeldovich
approximation
is always closest to the exact solution.

The detailed expression for $M^2$ in terms of Fourier components of the
initial gravitational potential ($\phi_0(\vec r) = (2\pi )^{-3/2}
\int \, d^3k\phi_{\vec k}e^{i\vec k\vec r};\, \,
\langle \phi_{\vec k}\phi_{\vec k'}^{\ast }\rangle = \phi^2(k)\delta^{(3)}
(\vec k -\vec k')$ where $k=|\vec k|$ and $\delta^{(3)}$ is now the $3D$
delta-function) is quite complicated (see appendix):

\begin{eqnarray}
M^2 = {1\over 32\pi^4} \int_0^{\infty} k_1\phi^2(k_1) \, dk_1
\int_0^{\infty} k_2^3\phi^2(k_2)\,
\nonumber \\
 \left(2k_1k_2(k_1^4+4k_1^2k_2^2-
k_2^4)+(k_2^2-k_1^2)^3\ln {k_1+k_2 \over |k_1-k_2|}\right) dk_2 \,
\end{eqnarray}
but if the physical wavelengh $2\pi a/k$ making the main contribution to
$\sigma_1$ is much larger than corresponding lengths for $\sigma_2$ and
$\sigma_3$, then the integrals in Eq. (24) decouple and $M^2 \approx
{1\over 3}\sigma_1^2 \sigma_2^2$. In this case, the Zeldovich approximation
is much closer to the exact solution than the other two. For a power-law
spectrum, this happens if $-3 < n \le -1$ (certainly, a cut-off at both some
large and small scales is implicitly assumed). On the other hand,
$M^2= \sigma_1^2 \sigma_2^2 $ in the opposite case of a $\delta$-like
isotropic power spectrum, so that in this case the  relative accuracy of  the
approximations
is $0.27;\, 1.0;\, 0.53 $ for ZA, FF and LP respectively.

Juszkiewicz et al. have recently suggested that
moments of the dimensionless velocity divergence $\theta = \dot a^{-1}
div \vec u = -(3t/2a^2)\triangle V$ (chosen so that $\langle \theta^{(1)2}
\rangle = \langle \delta^{(1)2} \rangle$) may be useful statistical quantities
of study in the weakly nonlinear regime (Juszkiewicz et al. 1993).
The value of the skewness
parameter $T_3 = \langle \theta^3 \rangle /(\langle \theta^2 \rangle)^2$
for $\theta$ can be determined from the results obtained in the previous
section. We find that in second order $T_3 =  3\langle (\theta^{(1)})^2
\theta^{(2)}\rangle/\langle (\theta^{(1)})^2\rangle^2 =$
 $(-{26\over 7}\approx -3.71,
{}~-2,~0,~-0.8)$
for the exact solution
and for the Zeldovich, frozen flow and linear potential approximations
respectively. We find that once more the ZA is the most accurate of the three
approximations, although the accuracy
of all approximations  worsens in this case.

One can also calculate the fractional error $F_\theta$ for $\theta$
just as we had done for $\delta$ earlier.
We define  the difference $ D_Z =\theta_Z^{(2)} -\theta^{(2)} $ (similarly for
$D_f$ and $D_p$) and construct fractional
quantities   $F_\theta = \langle{D_i}^2\rangle/
\langle \theta^{(2)2} \rangle$  ($i = Z, f, p$).
As in the case of  $F_\delta$,  $F_\theta$ also has the general form  $F_\theta
= (\bar a\gamma^2+\bar b)/
(\bar c\gamma^2+1)$. Where the coefficients $\bar a, \bar b,
\bar c$ have the values  (0.59, 0.0, 3.47) for ZA, (0.0, 1.0, 0.0) for  FF,
and (2.83, 0.75, 3.47) for LP respectively.
 $F_\theta$ is shown as a function of $\gamma$ for the
different approximations in figure 2. It is clear that as found earlier in
the case of
density perturbations, the Zeldovich approximation performs better than the
other two approximations for all values of $\gamma$.

\medskip

It is also possible to calculate cross correlations between the two fields
$\delta$ and $\theta$. Using results derived earlier one can easily
show that in second order $X_{12}=\langle \delta \theta^2 \rangle/
\sigma_{\rho}^4 = {1\over 3}(S_3-2T_3)$ and  $X_{21}=\langle
\delta^2 \theta \rangle/\sigma_{\rho}^4 = {1\over 3}(T_3-2S_3)$.

\section{Discussion}
Our results show that the Zeldovich approximation which is the
simplest and computationally the most cost-effective of the three
approximation methods considered by us in this paper,
is also more accurate {\it on an average} than
either the frozen flow or the linear potential approximation when
studied to second order in perturbation theory.
This might suggest that the Zeldovich approximation is a better
tool than either of the other two approximations with which
to study overdensities in the weakly nonlinear regime and also to
probe the dynamics of underdense regions such as voids.
Efforts to compare the different nonlinear approximation methods
discussed in the present paper with N-body simulations in the strongly
nonlinear regime are presently in progress (Sathyaprakash et al. 1993).

In addition, the Zeldovich approximation appears to be much closer to
exact density and velocity distributions than the other approximations
if the slope of the density power spectrum is $-3< n \le -1$ (just the range
for which we expect the prolonged existence of a network structure
qualitatively described by the adhesion model {\it after} caustic formation).
This also means that the deviation of the Zeldovich solution from the exact one
in the second order is much less on an average than the second-order term
itself.
This result may be generalized to higher orders of perturbation theory in
Eulerian space as well, i.e., $ \langle \Delta_Z^{(n)2}\rangle \ll \langle
\delta^{(n)2}
\rangle $ and $\langle \vec U_Z^{(n)2}\rangle \ll \langle \vec u^{(n)2}
\rangle $. However, this does not mean that the Zeldovich approximation
exactly describes a non-linear evolution even in this case because this
closeness originates from the fact that the effect of point displacement
(accurately taken into account by the Zeldovich approximation) is more
important for averaged values than effects of non-linearity for this range of
slopes.

Indeed, leading terms in all orders of perturbation theory having the largest
power of a large-scale velocity may be summed with the result (the same one for
both the exact solution and the Zeldovich approximation, see also Shandarin
1993):

\begin{eqnarray}
\delta = \delta^{(1)}(\vec q, t);~~~~~~~~~~~~~~ \nonumber \\
 \vec u =\vec u^{(1)}(\vec q, t);~~~~~~~~~~~~~~ \nonumber\\
\Phi = {V\over t}= \phi_0(\vec q)-{3t^2\over 4a^2}\left(\nabla
\phi_0(\vec r)\right)^2 ; \nonumber\\
 \vec q = \vec r +{3t^2\over 2a^2}\nabla \phi_0(\vec r).~~~~~~~~~~
\end{eqnarray}
This quasi-linear solution which is actually the linear
solution in Lagrangian space correctly accounts for a shift of phases
produced by the long-wave part of the perturbation spectrum but does not
describe genuinly non-linear effects which are mainly due to the short-wave
part (the substitution of $\vec r$ by $\vec q$ in the last terms of the
expressions for $\Phi $ and $\vec q$ that would exactly reproduce the Zeldovich
approximation for these quantities exceeds the accuracy with which Eq. (25)
is derived). That is why,
for instance, the error of the Zeldovich approximation in determining $S_3$ is
not too small.

 In a companion paper (Munshi et.al 1993) we obtained values of
higher moments of the distributions (i.e. $S_4, S_5,$~.. and $T_4, T_5,$~..)
as well. Then this are used to obtain $P(\delta)$ vs $\delta$,
$\delta$ vs $\theta$ relations and
void probability distribution function and related quantities.
\acknowledgements
The authors are grateful to V. Sahni for numerous stimulating discussions, to
D.Ju. Pogosyan and B.S. Sathyaprakash for comments on the paper and to
J.S. Bagla and  T. Padmanabhan and also to F.R. Bouchet for communicating their
papers prior to
publication. A.S. is grateful to Professors N. Dadhich and J.V. Narlikar for
their hospitality during his stay
in  IUCAA where this research was started and to the International
Science Foundation for supplying him with a travel grant for
participation in
the 6th Asian-Pacific Regional Meeting of the IAU held in IUCAA.
The financial support for the research work of A.S. in Russia was provided by
the Russian Foundation for Fundamental Research, Project Code 93-02-3631.
D.M. was financially supported by the Council of Scientific and Industrial
Research, India under its JRF scheme.

\centerline{ \bf APPENDIX}
\centerline {\bf DERIVATION OF EXPRESSION FOR ${\bf M^2}$}
\medskip
\medskip

We  derive the formula for $M^2$ used in section 3.2. By definition $M^2 =
 - \langle P\triangle^{-1} P\rangle $ It is useful to perform the analysis
in k space. Decomposing the gravitational potential in k space we get
$$\phi_0(\vec r)={1\over (2\pi )^{3/2}} \int \phi (\vec k)
e^{i\vec k\vec r}\, d^3k;\\ \eqno(A1)$$
$$P=\nabla (\nabla \phi_0 \triangle \phi_0)=
-\nabla {1\over (2\pi )^3} \int d^3k_1 \int d^3k_2 \, i\vec k_1k_2^2
\phi (\vec k_1) \phi (\vec k_2)e^{i(\vec k_1 + \vec k_2)\vec r} \nonumber\\ $$
$$={1\over (2\pi )^3}\int d^3k_1 \int d^3k_2\, \vec k_1(\vec k_1+
\vec k_2) k_2^2\phi(\vec k_1) \phi(\vec k_2)e^{i(\vec k_1+\vec k_2)\vec r}; \\
\eqno(A2)$$
$$\triangle^{-1} P = -{1\over (2\pi )^3} \int d^3k_1 \int d^3k_2 \,
{\vec k_1(\vec k_1+\vec k_2)k_2^2 \over (\vec k_1+\vec k_2)^2}
\phi (\vec k_1) \phi (\vec k_2)e^{i(\vec k_1+\vec k_2)\vec r}; \\ \eqno(A3)$$
$$\nabla (\triangle^{-1} P) = -{i\over (2\pi )^3} \int d^3k_1 \int d^3k_2
{(k_1^2+\vec k_1 \vec k_2)(\vec k_1+\vec k_2)k_2^2 \over (\vec k_1+\vec k_2)^2}
\phi (\vec k_1) \phi (\vec k_2)e^{i(\vec k_1+\vec k_2)\vec r}; \\ \eqno(A4)$$
$$M^2={1\over (2\pi )^6}\int d^3k_1 \int d^3k_2 \int d^3k_3 \int d^3k_4 \,
e^{i(\vec k_1+\vec k_2)\vec r - i(\vec k_3+ \vec k_4)\vec r} \nonumber \\ $$
$$\cdot {(k_1^2+\vec k_1 \vec k_2)(k_3^2+\vec k_3 \vec k_4)k_2^2k_4^2
\left( (\vec k_1+\vec k_2)(\vec k_3+\vec k_4)\right) \over
(\vec k_1+\vec k_2)^2(\vec k_3+\vec k_4)^2}
\langle \phi (\vec k_1) \phi (\vec k_2) \phi^{\ast} (\vec k_3) \phi^{\ast}
(\vec k_4)\rangle \,.\eqno(A5)$$

The only non-zero average values are:
$$\langle \phi (\vec k_1) \phi^{\ast} (\vec k_2) \rangle =\delta^3 (\vec k_1
-\vec k_2) \phi^2(k_1); \, \langle \phi (\vec k_1) \phi (\vec k_2)\rangle=
\delta^3 (\vec k_1+\vec k_2) \phi^2(k_1),\eqno(A6)$$
the last expression follows from the reality condition $\phi (-\vec k)=
\phi^{\ast} (\vec k)$. So,
$$M^2={1\over (2\pi )^6}\int d^3k_1 \int  \, \phi^2(k_1) \phi^2(k_2)$$
$$\left({(k_1^2+\vec k_1 \vec k_2)^2k_2^4\over (\vec k_1+\vec k_2)^2}
 +{(k_1^2+\vec k_1 \vec k_2)(k_2^2+\vec k_1 \vec k_2)k_1^2k_2^2\over
(\vec k_1+\vec k_2)^2}\right)   d^3k_2\\$$
$$= {1\over (2\pi )^6}\int d^3k_1 \int d^3k_2 \, \phi^2(k_1) \phi^2(k_2)
{(k_1^2+\vec k_1 \vec k_2)k_2^2 \over (\vec k_1+\vec k_2)^2}
(2k_1^2k_2^2+\vec k_1 \vec k_2 (k_1^2+k_2^2))  \nonumber  \\$$
$$={4\pi \cdot 2\pi \over (2\pi )^6}\int_0^{\infty} k_1^4 \phi^2(k_1)dk_1
\int_0^{\infty} k_2^5 \phi^2(k_2) dk_2 \int_{-1}^1 dz\, {(k_1+k_2z)
(2k_1k_2+z(k_1^2+k_2^2))\over k_1^2+k_2^2+2k_1k_2z}\,,\eqno(A7) $$
where z is the cosine of the angle between $\vec k_1$ and $\vec k_2$.
Performing the $z$ integration using standard textbook formulas one gets the
expression for $M^2$
$$M^2 = {1\over 32\pi^4} \int_0^{\infty} k_1\phi^2(k_1) \, dk_1
\int_0^{\infty} k_2^3\phi^2(k_2)\,
\nonumber \\$$
$$ \left(2k_1k_2(k_1^4+4k_1^2k_2^2-
k_2^4)+(k_2^2-k_1^2)^3\ln {k_1+k_2 \over |k_1-k_2|}\right) dk_2 \,\eqno(A8)$$
which we have used in the text.

\clearpage
{\bf Figure Captions}

{\bf Fig. 1} : The fractional error $F_{\delta} = \langle
\Delta^2\rangle/\langle
\delta^{(2)2} \rangle$ in $\delta$ for different approximations is shown as a
function of
$\gamma= \sigma^2_2/{\sigma_1\sigma_3}$. The solid, dashed, and dotted lines
correspond to the Zeldovich, frozen flow and linear potential approximations
respectively.
\medskip

{\bf Fig. 2} : The Fractional error $F_{\theta} = \langle D^2\rangle/\langle
\theta^{(2)2} \rangle$ in $\theta$ for different approximations is shown as a
function of
$\gamma= \sigma^2_2/{\sigma_1\sigma_3}$.
The solid, dashed, and dotted lines correspond to the Zeldovich, frozen flow
and linear potential approximations respectively.


\clearpage

\begin {thebibliography}{}
\bibitem{} Bagla, J.S., \& Padmanabhan, T., 1993. {\it MNRAS}, in press.  \\
\bibitem{} Bouchet, F.R, Juszkiewicz, R., Colombi, S., Pellat, R., 1992. {\it
ApJ} ~
{\bf 394}, L5.\\
\bibitem{} Brainerd, T.G., Scherer, R.J., \& Villumsen, J.V., 1992. {\it
Preprint
OSA-TA-12/92.} \\
\bibitem{} Grinstein, B., \& Wise, M.B., 1987. {\it ApJ}, ~{\bf 320}, 448.\\
\bibitem{} Gurbatov, S.N., Saichev, A.I., \& Shandarin, S.F., 1985. {\it Soviet
Phys.Dokl.}, {\bf 30}, 921.  \\
\bibitem{} Juszkiewicz, R., Weinberg, D.H., Amsterdamsky, P., Chodorovski, M.,
and
\bibitem{} Bouchet, F.R., 1993. {\it IAS preprint (IASSNS-AST 93/50)}.\\
\bibitem{} Kofman, L.A., 1991, in: {\it Primordial Nucleosynthesis and
Evolution   of Early Universe, eds. K.Sato} \& {\it J. Audoze } (Dordrecht :
Kluwer), p 495. \\
\bibitem{} Matarrese, S., Lucchin, F., Moscardini, L., \& Saez, D., 1992.
{\it MNRAS}, {\bf 259}, 437. \\
\bibitem{}  Munshi, D., Sahni, V., \& Starobinsky, A.A., 1993, in preparation.
\\
\bibitem{} Peebles, P.J.E., 1980. {\it The Large-Scale Structure of the
Universe},
Princeton, Princeton University Press. \\
\bibitem{} Sathyaprakash, B.S., Munshi, D., Sahni, V., Pogosyan, D. \& Melott,
A.L.,
1993, in preparation.\\
\bibitem{} Shandarin, S.F., 1993. {\it Kansas University preprint}; in: {\it
Proc. of the
dedication seminar for the Devayni Complex of Institutional Buildings of IUCAA,
Puna, India, Dec. 29-30, 1992.}  \\
\bibitem{} Shandarin, S.F., \& Zeldovich, Ya. B., 1989., {\it Rev. Mod. Phys.},
{\bf 61},185.
\bibitem{} Zeldovich, Ya.B., 1970. {\it Astron.Astroph.}, {\bf 5}, 84.

\end{thebibliography}
\end{document}